%% file: nips_2016.tex
\documentclass{article}

\PassOptionsToPackage{numbers}{natbib}
\usepackage[final]{nips_2016}

\usepackage[isolatin]{inputenc} 
\usepackage[T1]{fontenc}    
\usepackage{hyperref}       
\usepackage{url}            
\usepackage{booktabs}       
\usepackage{amsfonts}       
\usepackage{nicefrac}       
\usepackage{microtype}      

\usepackage{amsfonts,amsmath,amssymb,wasysym} 
\usepackage{graphicx,psfrag,multirow}
\usepackage{subfig}
\usepackage{algorithm}
\usepackage{algpseudocode}
\usepackage{wrapfig}
\usepackage{multirow}
\usepackage{color}
\usepackage{appendix}

\title{Approximate Recursive Identification of\\Autoregressive Systems with Skewed Innovations}

\author{
  Henri Nurminen\\
  Dept.\ of Automation Science and Engineering\\
  Tampere University of Technology\\
  Tampere, Finland\\
  \texttt{henri.nurminen@tut.fi} \\
 \And
 Tohid Ardeshiri\\
 Department of Engineering\\
 University of Cambridge\\
 Cambridge, UK\\
 \texttt{ta417@cam.ac.uk}\\
}

\begin{document}

\maketitle

\begin{abstract}
We propose a novel recursive system identification algorithm for linear autoregressive systems with skewed innovations. The algorithm is based on the variational Bayes approximation of the model with a multivariate normal prior for the model coefficients, multivariate skew-normally distributed innovations, and matrix-variate-normal--inverse-Wishart prior for the parameters of the innovation distribution. The proposed algorithm simultaneously estimates the model coefficients as well as the parameters of the innovation distribution, which are both allowed to be slowly time-varying. Through computer simulations, we compare the proposed method with a variational algorithm based on the normally-distributed innovations model, and show that modelling the skewness can provide improvement in identification accuracy.
\end{abstract}

\input{body}
\pagebreak
\subsection*{Acknowledgments}
H.\ Nurminen receives funding from Tampere University of Technology Graduate School, Nokia Technologies Oy, the Foundation of Nokia Corporation, and Tekniikan edist\"amiss\"a\"ati\"o.

T.\ Ardeshiri receives funding from Swedish research council's (VR) project Scalable Kalman filters, and from Jaguar Land Rover (JLR), Whitley, UK.

\medskip
\small
\bibliographystyle{ieeetr}
\bibliography{nips_2016}
\clearpage
\normalsize

\appendixpage
\appendix
\input{derivations}
\clearpage
\input{algo}
\end{document}

%% file: body.tex

\newcommand{\todo}[1]{\textcolor{red}{#1}}

\newcommand{\eye}{I}
\newcommand{\ones}{\mathbf{1}}
\newcommand{\zeros}{\mathrm{O}}
\newcommand{\ud}{\mathop{}\!\mathrm{d}}
\renewcommand{\d}{\mathop{}\!\mathrm{d}}
\newcommand{\N}{\mathrm{N}}
\newcommand{\MN}{\mathrm{N}}
\newcommand{\SN}{\mathrm{SN}}
\newcommand{\IW}{\mathrm{IW}}
\newcommand{\UNI}{\mathrm{unif}}
\newcommand{\R}{\mathbb{R}}
\newcommand{\diag}[1]{\mathrm{diag}(#1)}
\newcommand{\kron}{\otimes}
\newcommand{\KLD}{\mathrm{D}_{\mathrm{KL}}}
\newcommand{\E}{\mathbb{E}}
\renewcommand{\t}{\mathrm{T}}
\newcommand{\Tr}{\mathrm{Tr}}
\newcommand{\var}{\mathbb{V}}
\newcommand{\definedby}{\triangleq}
\newcommand{\nar}{n_\text{AR}}
\newcommand{\my}{y^{\ast}}
\newcommand{\cmy}{Y^{\ast}}
\newcommand{\oy}{y} 
\newcommand{\ay}{z} 

\section{Introduction}

Many systems produce datasets with skewed noise distribution. Skewness means asymmetry. Positive skewness, for example, intuitively means producing large positive deviations from the median value more frequently than large negative deviations. For instance, some financial data sets show negative skewness because large drops tend to be more frequent than large upsurges \cite{harvey1999,jondeau2003,christofferssen2006,tsiotas2012}. Wireless network based positioning often uses time delay measurement as a distance, but non-line-of-sight can produce large positive outliers, so the error distribution becomes positively skewed \cite{kaemarungsi2012,nurminen2015b}.



One statistical model for skewed error distributions is the skew normal distribution and its multivariate generalisation \cite{azzalini1985, azzalini1996}. The posterior distribution of a normal prior and skew-normal measurement noise model is not analytically tractable. However, the distribution admits a hierarchical formulation whose favorable conjugacy properties enable efficient parameter estimation using the expectation--maximisation (EM) algorithm \cite{lin2009,lee2014,ho2012} and approximate Bayesian time-series filtering and smoothing based on the variational Bayes (VB) approximation \cite{nurminen2015a,nurminen2016b_arxiv}.

This paper studies autoregressive (AR) models, where the measurement is modelled to be a linear function of $\nar$ (the model order) previous measurements plus an independent random noise term referred to as the innovation. When the AR coefficients and/or the conditionally skew-normal innovation distribution's statistics are time-varying or they need to be identified online, recursive identification methods are used \cite{ljung2002}.

In this paper we propose a novel recursive system identification algorithm for AR models with skew-normally distributed measurement noise with unknown possibly slowly time-varying scale and skewness. The proposed approximation is based on a VB approximation.

\section{Problem formulation}

\begin{figure}[ht]
\centering
\vspace{-2mm}
\includegraphics[width=0.6\textwidth]{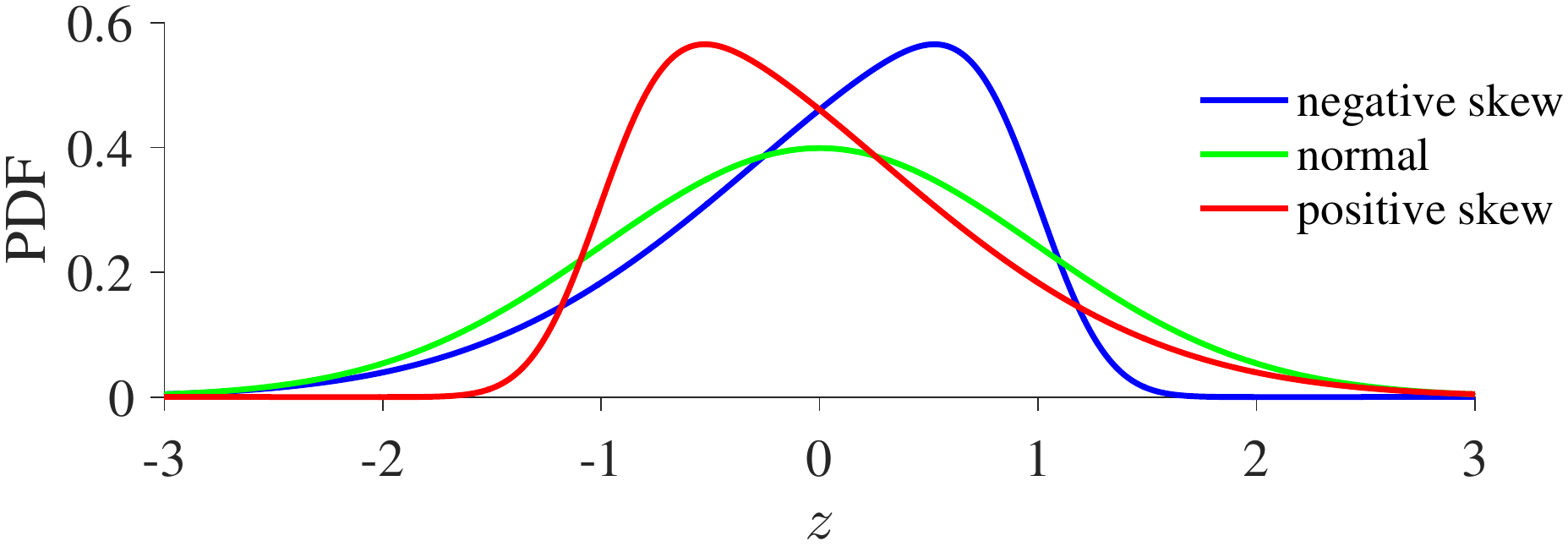}
\vspace{-2mm}
\caption{The PDFs of negatively-skewed, symmetric, and positively-skewed normal distributions. Each distribution has mean zero and variance one.} \label{fig:pdfs}
\end{figure}

Skew normal distribution is an asymmetric generalization of the normal distribution originally proposed by Azzalini \cite{azzalini1985}. Its multivariate version was later introduced by Azzalini and Dalla Valle \cite{azzalini1996}. The version that is used in this report is the canonical fundamental skew normal distribution (CFUSN) introduced by Arellano Valle and Genton \cite{arellanovalle2005}. However, we adopt a different parametrization following the guidelines of the canonical fundamental skew $t$-distribution's parametrization in \cite{lee2016} to obtain a suitable analytical tractability. The probability density function (PDF) of this skew normal distribution $z \sim \SN(\mu,R,\Delta)$ is
\begin{align}
p(z) &= 2^{n_z} \, \N(z; \mu-\Delta\sqrt{\tfrac{2}{\pi}}\ones, \Omega) \, F_{\N}(\Delta^\t \Omega^{-1} (z-\mu+\Delta\sqrt{\tfrac{2}{\pi}}\ones); 0, \eye-\Delta^\t \Omega^{-1} \Delta) ,
\end{align}
where $\ones$ is a vector of ones, $\mu$ is a location parameter, $\Omega=R+\Delta\Delta^\t$, and $F_{\N}$ is the cumulative distribution function of the multivariate normal distribution. $R\in\R^{n_z \times n_z}$ (symmetric positive-definite, spd) and $\Delta\in\R^{n_z \times n_z}$ are shape matrices that determine the scale and skewness, and the sign and structure of $\Delta$ determine the direction of skewness as explained in \cite{lee2016}. Examples of the PDF in negatively skewed, symmetric, and positively skewed cases are given in Fig.\ \ref{fig:pdfs}. The moments of this multivariate skew normal distribution given the shape matrices are
\begin{align}
\E[z] = \mu,\quad \var[z] = R + \tfrac{2}{\pi} \Delta\Delta^\t .
\end{align}
Compared to the formulation of \cite{lee2016}, we shift the distribution with $\Delta\sqrt{2/\pi}\ones$ so that the mean of the distribution does not depend on $\Delta$. This ensures that the proposed algorithm identifies $\Delta$ as a measure of skewness, not as a measure of location. 

We formulate the AR coefficient estimation problem as the linear state-space model with the measurement noise being skew-normally distributed conditional on the unknown slowly-varying noise parameters $R_k$ and $\Delta_k$
\begin{subequations} \label{eq:ssm}
\begin{align}
&p(x_1) = \N(x_1;x_{1|0},P_{1|0}) \\
&x_k = x_{k-1} + w_{k-1},\quad w_{k-1} \stackrel{iid}{\sim} \N(0,Q_{k-1}) \label{eq:state_transition} \\
&z_k = C_k x_k + e_k,\quad e_k \stackrel{iid}{\sim} \SN(\mu,R_k,\Delta_k) ,
\end{align}
\end{subequations}
where $x_k \in\R^{\nar}$ is the vector of AR coefficients, $Q_k \in \R^{\nar \times \nar}$ (spd) is the process noise covariance matrix that is assumed known and is thus an algorithm parameter, $z_k \in \R^{n_z}$ is the measurement, $C_k \in \R^{n_z \times \nar} \!=\! \left[\begin{smallmatrix} z_{k-1}&z_{k-2}&\cdots&z_{k-\nar} \end{smallmatrix}\right]$ is the measurement model matrix given by $\nar$ previous measurements, and $\{w_k\in\R^{\nar}\}_{k=1}^K$ and $\{e_k\in\R^{n_z}\}_{k=1}^K$ are mutually independent process and measurement noise sequences.

\section{Proposed algorithm}

\subsection{Measurement update}

Conditional on the parameters $R_k$ and $\Delta_k$, the skew-normal random variable $e_k|R_k,\Delta_k\sim\SN(\mu,R_k,\Delta_k)$ has the hierarchical formulation \cite{lin2009}
\begin{subequations}
\begin{align}
e_k|u_k,R_k,\Delta_k &\sim \N(\mu + \Delta_k (u_k-\sqrt{\tfrac{2}{\pi}}\ones), R_k) \\
u_k &\sim \N_+(0,\eye) ,
\end{align}
\end{subequations}
where $\N_+$ is the multivariate normal distribution truncated into positive orthant. To obtain the necessary conjugacy properties, let us assign the matrix-variate-normal--inverse-Wishart (MVNIW) prior distribution to the joint random variable $(R_k,\Delta_k)$:
\begin{align}
p(R_k,\Delta_k) &= \MN(\Delta_k ; \Delta_{k|k-1}, R_k \kron V_{k|k-1})\, \IW(R_k; \Psi_{k|k-1},\nu_{k|k-1}) , \label{eq:RD_prior}
\end{align}
where $\Delta_{k|k-1} \in \R^{n_z \times n_z}$, $V_{k|k-1} \in \R^{n_z \times n_z}$ (spd), $\Psi_{k|k-1} \in \R^{n_z \times n_z}$ (spd), and $\nu_{k|k-1} \in (2n_z,\infty)$ are parameters of the prior distribution. $\MN(X; M,U \kron V)$ is the PDF of the matrix-variate normal distribution with mean $M$, and variance parameters $U$ (among-row) and $V$ (among-column) \cite[Ch.\ 2]{GuptaN:2000}, and $\IW(X; \Psi,\nu)$ is PDF of the inverse-Wishart distribution with scale-matrix $\Psi$ and $\nu$ degrees of freedom \cite[Ch.\ 3]{GuptaN:2000}.

The filtering posterior distribution $p(x_k,u_k,R_k,\Delta_k|\ay_{1:k})$ of the model defined by \eqref{eq:ssm} and \eqref{eq:RD_prior} is not analytically tractable. Our solution is to use a variational Bayesian approximation, where we find the functions $q_{x,u}(x_k,u_k)$ and $q_{R,\Delta}(R_k,\Delta_k)$ such that the reversed Kullback--Leibler divergence (KLD)
\begin{equation} \label{eq:kld}
\KLD\big(q_{x,u}(x_k,u_k)\,q_{R,\Delta}(R_k,\Delta_k) \| p(x_k,u_k,R_k,\Delta_k|\ay_{1:k})\big)
\end{equation}
is minimised, where $\KLD(q\|p) \!=\! \int q(x) \log(\tfrac{q(x)}{p(x)}) \ud x$. In general there is no exact analytical solution for $({q}_{x,u},{q}_{R,\Delta})$, but the iteration of
\begin{subequations}
\label{eq:FilterIterativeOptimization}
\begin{align}
&\log {q}_{x,u}(x_{k},u_k) \leftarrow \E_{q_{R,\Delta},}[\log p(\ay_k,x_{k},u_{k},R_k,\Delta_k|\ay_{1:k-1})]+c_{x,u}\label{eq:FilterIterativeOptimizationx}\\
&\log {q}_{R,\Delta}(R_k,\Delta_k) \leftarrow \E_{{q}_{x,u}}[\log p(\ay_k,x_{k},u_{k},R_k,\Delta_{k}|\ay_{1:k-1})]+c_{R,\Delta}\label{eq:FilterIterativeOptimizationDR}
\end{align}
\end{subequations}
always reduces the KLD \eqref{eq:kld} and for many models gives a sequence that converges towards the optimal functions $({q}_{x,u},{q}_{R,\Delta})$~\cite[Chapter 10]{Bishop2007}\cite{TzikasLG2008}. The expected values on the right hand sides of~\eqref{eq:FilterIterativeOptimization} are taken with respect to the current $q_{x,u}$ and $q_{R,\Delta}$, and $c_{x,u}$ and $c_{R,\Delta}$  are constants with respect to the variables $(x_k,u_k)$, and $(R,\Delta)$, respectively. 

Thanks to the chosen prior distribution structure \eqref{eq:RD_prior}, the update \eqref{eq:FilterIterativeOptimizationDR} has a closed form solution that preserves the functional form of the prior, and the moments of the distribution required by other computations are also analytically tractable. The analytical solution of the update \eqref{eq:FilterIterativeOptimizationx} is a multivariate normal distribution truncated by multiple linear constraints. The mean and covariance matrix of this distribution can be approximated using the sequential truncation algorithm \cite{perala2008, simon2010, nurminen2016b_arxiv}. The distribution $q_{x,u}(x_k,u_k)$ is then approximated by the unconstrained multivariate normal distribution with the obtained moments
\begin{align}
q_{x,u}(x_k,u_k) &\approx \N\left( \left[\begin{smallmatrix} x_k\\u_k \end{smallmatrix}\right]; \xi_{k|k}, \Xi_{k|k} \right),
\end{align}
where $\xi_{k|k}$ and $\Xi_{k|k}$ are the mean and covariance matrix given by the sequential truncation algorithm. Normal marginal posterior approximation for $x_k$ guarantees that we get a recursive algorithm. The approximative filtering posterior of $(R_k,\Delta_k)$ is the MVNIW distribution
\begin{align} \label{eq:qRD}
q_{R,\Delta}(R_k,\Delta_k) &= \MN(\Delta_k; \Delta_{k|k}, R_k \kron V_{k|k}) \, \IW(R_k; \Psi_{k|k},\nu_{k|k}) ,
\end{align}
whose required moments are analytically tractable when $\nu_{k|k}\!>\!2n_z$ as shown in Appendix \ref{app:derivations}.

\subsection{Time update} \label{sec:time_update}

The marginal distribution of the AR coefficient vector $x_k$ in the posterior approximation $\N\left( \left[\begin{smallmatrix} x_k\\u_k \end{smallmatrix}\right]; \xi_{k|k}, \Xi_{k|k} \right) \cdot q_{R,\Delta}(R_k,\Delta_k)$ is a normal distribution and the state transition \eqref{eq:state_transition} is linear and Gaussian. Thus, the filter prediction becomes the standard Kalman filter prediction and the prediction distribution is normal.

The dynamical model of the model parameters $p(R_{k},\Delta_{k} | R_{k-1},\Delta_{k-1})$ is typically unknown and/or intractable. Therefore, we adopt the forgetting factor update, which provides the maximum-entropy solution given the KLD from the previous posterior \cite{karny2012, ozkan2013}. Thus, the used prediction density given the MVNIW approximation of the previous posterior and the forgetting factor $\gamma \in (0,1]$ is
\begin{align} \label{eq:time_update}
\hat{p}(R_{k},\Delta_{k}| y_{1:k-1}) &\propto \MN(\Delta_k; \Delta_{k-1|k-1},R_k \kron \tfrac{1}{\gamma} V_{k-1|k-1}) \nonumber\\
&\times \IW(R_k; \gamma \Psi_{k-1|k-1}, \gamma \,\nu_{k-1|k-1} + (1\!-\!\gamma)\!\cdot\! 2n_z).
\end{align}
where the term $(1\!-\!\gamma)\cdot 2n_z$ guarantees that the resulting inverse-Wishart distribution is well-defined and has an expectation value.

The details of the proposed recursive identification algorithm including the prediction equations implied by the time update \eqref{eq:time_update} are given in Appendix \ref{table:filter}.

\section{Simulated example}

We simulated 1000 Monte Carlo replications of the AR model with 25 AR coefficients with $n_z\!=\!2$ dimensional skew-normally distributed innovations with parameters $R\!=\!0.1^2 \!\cdot\! \eye$ and $\Delta\!=\!\left[ \begin{smallmatrix} 2&0\\1&2 \end{smallmatrix} \right]$. Thus, the true distribution has high positive skewness. The true coefficients were simulated by generating the zeros of the characteristic polynomial from the uniform distribution $\UNI(-1,1)$. 
The number of AR coefficients was assumed known. The initial prior covariance matrix for the AR coefficient vector was given by the 1st order stable spline kernel $[P_{1|0}]_{i,j}=\frac{30-1}{3} 0.5^{\max(i-1,j-1)}$, and the process noise covariance was chosen as $[Q_{k-1}]_{i,j}=(\tfrac{1}{\gamma}-1)\max(\diag{P_{k-1|k-1}}) \cdot 0.5^{\max(i-1,j-1)}$ to preserve the stable kernel form of the prior \cite{pillonetto2010,chen2012}.

\begin{figure}[h]
\centering
\includegraphics[width=0.7\textwidth]{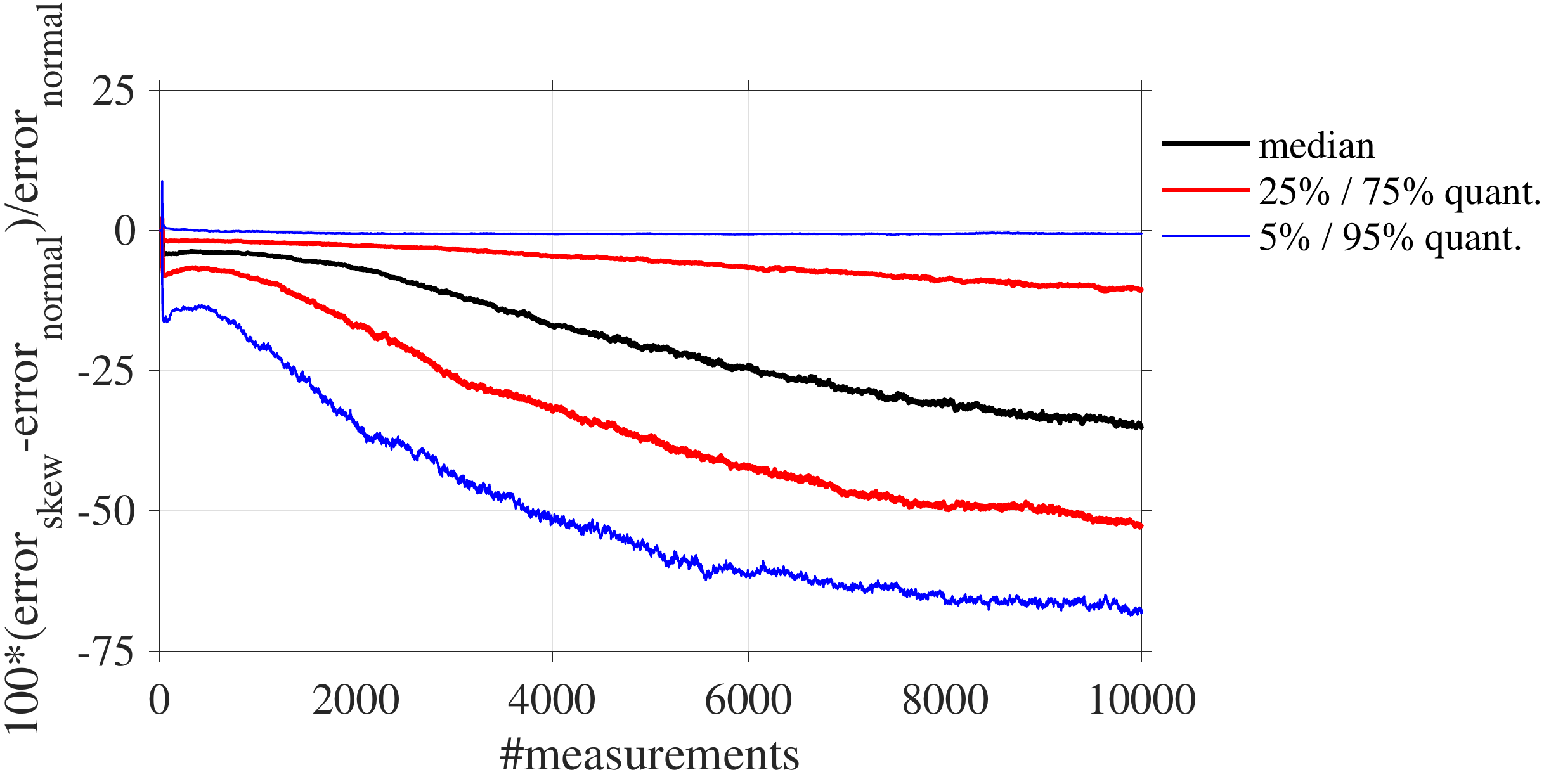}
\caption{The proposed algorithm is more accurate than the normal distribution based algorithm in 95\,\% of the simulations, and in most of the simulations the error \eqref{eq:id_error} is reduced by more than 25\,\%.} \label{fig:simulation_diff}
\end{figure}

The proposed method is compared with the Gaussian variational Bayes filter for slowly-drifting noise proposed by Agamennoni et al in \cite{Agamennoni12}. The skew-normal based identification method was  given the positive direction of the skewness by using the initial prior
\begin{align}
p(R_1,\Delta_1) = \N(\Delta_1; \sqrt{\tfrac{\pi}{2}\tfrac{1}{2}} \eye, R_1 \kron \eye) \, \IW(R_1; \tfrac{\nu_{1|0}\!-\!3}{2} \eye, \nu_{1|0}) ,
\end{align}
where $\nu_{1|0}\!=\!4\!+\!10^{-10}$. That is, the variance is divided equally between the symmetric and skewed component in the sense that $\E[R_1^{-1}]^{-1} = \E[\Delta_1]^2\tfrac{2}{\pi} = \tfrac{1}{2}\eye$. The normal distribution based method was given the initial prior
\vspace{-2mm}
\begin{align}
p(R_1) = \IW(R_1; (\nu_{1|0}\!-\!3) \eye, \nu_{1|0}) .
\end{align}
The forgetting factor value used with both the methods was $\gamma\!=\!0.975$, and the number of VB iterations was 10. Fig.\ \ref{fig:simulation_diff} shows the relative difference of the identification error
\begin{align} \label{eq:id_error}
\epsilon_k = \sqrt{ \textstyle{\sum_{i=1}^{\nar}} (x_{k|k}-(x_k)_\text{true})^2 }
\end{align}
as a function of the fed number of measurements. The figure shows that the skew-normal based identification method gives a lower median error than the normal distribution based, and the relative differences increase as the number of measurements increase. Fig.\ \ref{fig:simulation_diff} shows that after 10.000 measurements, the skew-normal based method is more accurate in about 95\,\% of the cases and gives at least 25\,\% lower identification error in most of the simulations.

\section{Conclusions}

We proposed a novel recursive estimation algorithm for identifying the model coefficients and innovation distribution parameters of autoregressive models with skew-normally distributed innovations. Both model coefficients and innovation distribution parameters can be slowly time-varying. Our computer simulation showed that modelling skewness can improve the accuracy of identification.

%% file: derivations.tex

\section{Variational solution of the measurement update} \label{app:derivations}

\allowdisplaybreaks[1]

Our variational solution uses this hierarchical formulation of the measurement noise model:
\begin{subequations}
\label{eq:hierarchical_adaptive}
\begin{align}
z_k |x_k,u_k,R_k,\Delta_k &\sim \N(C_k x_k+\Delta_k (u_k-\sqrt{\tfrac{2}{\pi}}\ones), R_k),\\
u_k &\sim \N_+(0, \eye),\\
\Delta_k|R_k &\sim \MN(\Delta_{k|k-1},R_k \kron V_{k|k-1}) , \\
R_k & \sim \IW(\Psi_{k|k-1},\nu_{k|k-1}) ,
\end{align}
\end{subequations}
where $z_k \in \mathbb{R}^{n_z}$ is the measurement, $u_k \in \mathbb{R}^{n_z}$ is the skewness variable vector, and $\Delta_{k|k-1} \in \R^{n_z \times n_z}$, $V_{k|k-1} \in \R^{n_z \times n_z}$ (spd), $\Psi_{k|k-1} \in \R^{n_z \times n_z}$ (spd), and $\nu_{k|k-1}>2n_z$ are the parameters of the joint prior distribution of $\Delta_k$ and $R_k$. The prior of $\Delta_k$ and $R_k$ is implied by the previous filtering posterior and the time update step (filter prediction) that is explained in section \ref{sec:time_update}.

The derivations for the variational solution \eqref{eq:FilterIterativeOptimization} are given in Sections \ref{sec:Filterq_x} and \ref{sec:Filterq_DR}. For brevity all constant values are denoted by $c$ in the derivation. The logarithm of the full filtering distribution which is needed for the derivations is\begin{align}
&\log p(z_k,x_k,u_k,R_k,\Delta_k |z_{1:k-1}) \nonumber\\
=& -\frac{1}{2}(\ay_k-C_k x_k-\Delta_k (u_k-\sqrt{\tfrac{2}{\pi}}\ones))^\t{}R_k^{-1} 
 (\ay_k-C_k x_k-\Delta_k (u_k-\sqrt{\tfrac{2}{\pi}}\ones))\nonumber\\
&-\frac{1}{2}(x_k-x_{k|k-1})^\t{}P_{k|k-1}^{-1}(x_k-x_{k|k-1}) - \frac{1}{2}u_k^\t{} u_k \nonumber\\
&-\frac{1}{2}\Tr\{(\Delta_k-\Delta_{k|k-1})V_{k|k-1}^{-1}(\Delta_k-\Delta_{k|k-1})^\t R_k^{-1} \} \nonumber\\
&-\frac{\nu_{k|k-1}+1}{2} \log\det(R_k) - \frac{1}{2} \Tr\{\Psi_{k|k-1}R_k^{-1}\} + c,\quad u\geq0,
\end{align}
where $x_{k|k-1}$ and $P_{k|k-1}$ are the mean and covariance matrix of the current predictive distribution, and $\Tr\{\cdot\}$ is the matrix trace.

\subsection{ Derivations for $q_{x,u}$}
\label{sec:Filterq_x}
Using equation \eqref{eq:FilterIterativeOptimizationx} we obtain
\begin{align}
&\log q_{x,u}(x_k,u_k) \nonumber\\
=&-\frac{1}{2}\E_{q_{R,\Delta}} \big[ (\ay_k-C_k x_k-\Delta_k (u_k-\sqrt{\tfrac{2}{\pi}}\ones))^\t R_k^{-1}
 (\ay_k-C_k x_k-\Delta_k (u_k-\sqrt{\tfrac{2}{\pi}}\ones)) \big]\nonumber\\
&-\frac{1}{2}(x_k-x_{k|k-1})^\t P_{k|k-1}^{-1}(x_k-x_{k|k-1}) - \frac{1}{2} u_k^\t u_k +c\\
=&- \frac{1}{2}(\ay_{k}-C_k x_k-\Delta_{k|k} (u_k-\sqrt{\tfrac{2}{\pi}}\ones))^\t R_{k|k}^{-1}(\ay_{k}-C_k x_k-\Delta_{k|k} (u_k-\sqrt{\tfrac{2}{\pi}}\ones))\nonumber\\
&-\frac{1}{2} (u_k-\sqrt{\tfrac{2}{\pi}}\ones)^\t (\E_{q_{R,\Delta}}[\Delta_k^\t R_k^{-1} \Delta_k] - \Delta_{k|k}^\t R_{k|k}^{-1} \Delta_{k|k}) (u_k-\sqrt{\tfrac{2}{\pi}}\ones) \nonumber\\
&-\frac{1}{2}(x_k-x_{k|k-1})^\t{}P_{k|k-1}^{-1}(x_k-x_{k|k-1}) - \frac{1}{2}u_k^\t u_k +c \quad u_k \geq 0
\label{eq:Filterqxposterior},
\end{align}
where $(R_{k|k},\Delta_{k|k}) \!\triangleq\! (\E_{q_{R,\Delta}}[R_k^{-1}]^{-1},\E_{q_{R,\Delta}}[\Delta_k])$ as well as the identity $\E_{q_{R,\Delta}}[ R_k^{-1} \Delta_k ] \!=\! R_{k|k}^{-1}\Delta_{k|k}$ are derived in Section \ref{sec:Filterq_DR}. The inequality $u_k\!\geq\!0$ denotes that each element of the vector $u_k$ is required to be greater or equal than zero. Further, in Section \ref{sec:Filterq_DR} it is proved that $\E_{q_{R,\Delta}}[ \Delta_k^\t R_k^{-1} \Delta_k ] = n_zV_{k|k}+\Delta_{k|k}^\t R_{k|k}^{-1} \Delta_{k|k}$, so Eq.\ \eqref{eq:Filterqxposterior} becomes
\begin{align}
&\log q_{x,u}(x_k,u_k) \nonumber\\
=&- \frac{1}{2}(\ay_{k}-C_k x_k-\Delta_{k|k} (u_k-\sqrt{\tfrac{2}{\pi}}\ones))^\t R_{k|k}^{-1}(\ay_{k}-C_k x_k-\Delta_{k|k} (u_k-\sqrt{\tfrac{2}{\pi}}\ones))\nonumber\\
&-\frac{n_z}{2} (u_k-\sqrt{\tfrac{2}{\pi}}\ones)^\t V_{k|k} (u_k-\sqrt{\tfrac{2}{\pi}}\ones) \nonumber\\
&-\frac{1}{2}(x_k-x_{k|k-1})^\t{}P_{k|k-1}^{-1}(x_k-x_{k|k-1}) - \frac{1}{2}u_k^\t u_k +c \\
=& -\frac{1}{2} (z_{k}+\Delta_{k|k}\sqrt{\tfrac{2}{\pi}}\ones-\left[ \begin{smallmatrix} C_k&\Delta_{k|k} \end{smallmatrix} \right] \left[\begin{smallmatrix} x_k\\u_k \end{smallmatrix}\right])^\t R_{k|k}^{-1} (z_{k}+\Delta_{k|k}\sqrt{\tfrac{2}{\pi}}\ones-\left[ \begin{smallmatrix} C_k&\Delta_{k|k} \end{smallmatrix} \right] \left[\begin{smallmatrix} x_k\\u_k \end{smallmatrix}\right]) \nonumber\\
&-\frac{1}{2} (\left[\begin{smallmatrix} x_k\\u_k \end{smallmatrix}\right]-\xi_{k|k-1})^\t \Xi_{k|k-1}^{-1} (\left[\begin{smallmatrix} x_k\\u_k \end{smallmatrix}\right]-\xi_{k|k-1}) + c, \quad u_k\geq0,
\end{align}
where
\begin{align}
\Xi_{k|k-1} &= \begin{bmatrix} P_{k|k-1} & \zeros \\ \zeros & (\eye+n_zV_{k|k})^{-1} \end{bmatrix}, \\
\xi_{k|k-1} &= \begin{bmatrix} x_{k|k-1} \\ n_z \sqrt{\tfrac{2}{\pi}}  (\eye+n_zV_{k|k})^{-1} V_{k|k} \ones \end{bmatrix} .
\end{align}

Hence, 
\begin{align}
q_{x,u}(x_k,u_k) &\propto  \N(\ay_{k}+\Delta_{k|k}\sqrt{\tfrac{2}{\pi}}\ones; \left[ \begin{smallmatrix} C_k&\Delta_{k|k} \end{smallmatrix} \right] \left[\begin{smallmatrix} x_k\\u_k \end{smallmatrix}\right], R_{k|k})\, \N( \left[\begin{smallmatrix} x_k\\u_k \end{smallmatrix}\right]; \xi_{k|k-1}, \Xi_{k|k-1} )\cdot [\![ u_k \geq 0 ]\!] \\
&\propto \N(\left[\begin{smallmatrix} x_k\\u_k \end{smallmatrix}\right]; \widehat\xi_{k|k}, \widehat\Xi_{k|k}) \cdot [\![u_k\geq 0]\!] ,
\end{align}
where $[\![ \cdot ]\!]$ is the Iverson bracket, and $\widehat\xi_{k|k}$ and $\widehat\Xi_{k|k}$ are the outputs of the Kalman filter update
\begin{align}
\widetilde{C}_k &= \left[ \begin{smallmatrix} C_k&\Delta_{k|k} \end{smallmatrix} \right] ,\\
K_k &= \Xi_{k|k-1} \widetilde{C}_k^\t (\widetilde{C}_k \Xi_{k|k-1} \widetilde{C}_k^\t + R_{k|k})^{-1},\\
\widehat\xi_{k|k} &= \xi_{k|k-1} + K_k (z_{k}+\Delta_{k|k}\sqrt{\tfrac{2}{\pi}}\ones - \widetilde{C}_k \xi_{k|k-1}),\\
\widehat\Xi_{k|k} &= (\eye-K_k \widetilde{C}_k) \Xi_{k|k-1} .
\end{align}
To make the algorithm recursive, we approximate $q_{x,u}$ with a multivariate normal distribution
\begin{align}
q_{x,u}(x_k,u_k) &= \N(\left[\begin{smallmatrix} x_k\\u_k \end{smallmatrix}\right]; \widehat\xi_{k|k}, \widehat\Xi_{k|k}) \cdot [\![u_k\geq 0]\!] \\
&\approx \N(\left[\begin{smallmatrix} x_k\\u_k \end{smallmatrix}\right]; \xi_{k|k}, \Xi_{k|k}) , \label{eq:appr_qxu}
\end{align}
whose approximate mean and covariance matrix $\xi_{k|k}$ and $\Xi_{k|k}$ are obtained through approximate moment-matching. Our approach for approximating the moments is the sequential truncation algorithm \cite{perala2008,simon2010}\cite[Table I]{nurminen2016b_arxiv}. Let us denote the approximate distribution with $\widetilde{q}_{x,u}(x_k,u_k) \triangleq \N(\left[\begin{smallmatrix} x_k\\u_k \end{smallmatrix}\right]; \xi_{k|k}, \Xi_{k|k})$.

In Section \ref{sec:Filterq_DR}, certain moments of $q_{x,u}$ are required. They are approximated as
\begin{align}
x_{k|k} &\triangleq \E_{\widetilde{q}_{xu}}[x_k] = [\xi_{k|k}]_{1:n_x}, \\
P_{k|k} &\triangleq \var_{\widetilde{q}_{xu}}[x_k] = [\Xi_{k|k}]_{1:n_x,1:n_x}, \\
u_{k|k} &\triangleq \E_{\widetilde{q}_{xu}}[u_k] = [\xi_{k|k}]_{n_x+(1:n_z)}, \\
U_{k|k} &\triangleq \var_{\widetilde{q}_{xu}}[u_k] = [\Xi_{k|k}]_{n_x+(1:n_z),n_x+(1:n_z)}, \\
\Upsilon_{k|k} &\triangleq \E_{\widetilde{q}_{xu}}[x_k u_k^\t]-x_{k|k}u_{k|k}^\t = [\Xi_{k|k}]_{1:n_x,n_x+(1:n_z)} ,
\end{align}
where $n_x\!+\!(1:n_z)$ denotes $(n_x\!+\!1):(n_x\!+\!n_z)$.

\subsection{ Derivations for $q_{R,\Delta}$}
\label{sec:Filterq_DR}
Using equation \eqref{eq:FilterIterativeOptimizationDR} and the approximation \eqref{eq:appr_qxu} we obtain
\begin{align}
\log &q_{R,\Delta}(R_k,\Delta_k) = \E_{\widetilde{q}_{x,u}}\left[\log \N(\ay_k;C_k x_k+\Delta_k (u_k-\sqrt{\tfrac{2}{\pi}}\ones), R_k) \right] \nonumber \\
& +\log \MN(\Delta_k; \Delta_{k|k-1}, R_k \kron V_{k|k-1}) + \log \IW(R_k; \Psi_{k|k-1},\nu_{k|k-1}) +c\\
=& -\frac{1}{2} \log\det(R_k) \nonumber\\
&-\frac{1}{2} \Tr \bigg\{ \E_{\widetilde{q}_{x,u}}\bigg[ (\ay_k-C_k x_k-\Delta_k (u_k-\sqrt{\tfrac{2}{\pi}}\ones)) (\ay_k-C_k x_k-\Delta_k (u_k-\sqrt{\tfrac{2}{\pi}}\ones))^\t \bigg] R_k^{-1} \bigg\} \nonumber\\
& -\frac{n_z}{2} \log\det(R_k) - \frac{1}{2} \Tr\left\{ (\Delta_k-\Delta_{k|k-1}) V_{k|k-1}^{-1} (\Delta_k-\Delta_{k|k-1})^\t R_k^{-1} \right\} \nonumber\\
& - \frac{\nu_{k|k-1}}{2} \log\det(R_k) -\frac{1}{2} \Tr\left\{ \Psi_{k|k-1} R_k^{-1} \right\} +c\\
=& - \frac{\nu_{k|k-1}+n_z+1}{2} \log\det(R_k) \nonumber\\
&- \frac{1}{2} \Tr \bigg\{ \bigg( \Delta_k \big( U_{k|k} + (u_{k|k}-\sqrt{\tfrac{2}{\pi}}\ones)(u_{k|k}-\sqrt{\tfrac{2}{\pi}}\ones)^\t \big) \Delta_k^\t \nonumber\\
&- \big(\ay_{k}(u_{k|k}-\sqrt{\tfrac{2}{\pi}}\ones)^\t -C_k (\Upsilon_{k|k}+x_{k|k}(u_{k|k}-\sqrt{\tfrac{2}{\pi}}\ones)^\t)\big) \Delta_k^\t  \nonumber\\
&- \Delta_k \big(\ay_{k}(u_{k|k}-\sqrt{\tfrac{2}{\pi}}\ones)^\t -C_k (\Upsilon_{k|k}+x_{k|k}(u_{k|k}-\sqrt{\tfrac{2}{\pi}}\ones)^\t)\big)^\t \nonumber\\
& + (\ay_{k}-C_k x_{k|k})(\ay_{k}-C_k x_{k|k})^\t + C_k P_{k|k}C_k^\t \bigg) R_k^{-1} \bigg\} \nonumber \\
&- \frac{1}{2} \Tr\left\{ \left( (\Delta_k-\Delta_{k|k-1}) V_{k|k-1}^{-1} (\Delta_k-\Delta_{k|k-1})^\t  + \Psi_{k|k-1} \right) R_k^{-1} \right\} + c \\
=& - \frac{\nu_{k|k-1}+n_z+1}{2} \log\det(R_k) \nonumber\\
&- \frac{1}{2} \Tr \bigg\{ \bigg( \Delta_k \big( U_{k|k} + (u_{k|k}-\sqrt{\tfrac{2}{\pi}}\ones)(u_{k|k}-\sqrt{\tfrac{2}{\pi}}\ones)^\t + V_{k|k-1}^{-1} \big) \Delta_k^\t \nonumber\\
&- \big(\ay_{k}(u_{k|k}-\sqrt{\tfrac{2}{\pi}}\ones)^\t -C_k(\Upsilon_{k|k}+x_{k|k}(u_{k|k}-\sqrt{\tfrac{2}{\pi}}\ones)^\t) + \Delta_{k|k-1} V_{k|k-1}^{-1} \big) \Delta_k^\t \nonumber\\
& - \Delta_k \big( \ay_{k}(u_{k|k}-\sqrt{\tfrac{2}{\pi}}\ones)^\t -C_k(\Upsilon_{k|k}+x_{k|k}(u_{k|k}-\sqrt{\tfrac{2}{\pi}}\ones)^\t) + \Delta_{k|k-1} V_{k|k-1}^{-1} \big)^\t \nonumber\\
&+ (\ay_{k}-C_k x_{k|k})(\ay_{k}-C_k x_{k|k})^\t + C_k P_{k|k}C_k^\t + \Delta_{k|k-1} V_{k|k-1}^{-1} \Delta_{k|k-1}^\t + \Psi_{k|k-1} \bigg) R_k^{-1} \bigg\}\\
=& -\frac{n_z}{2} \log\det(R_k) - \frac{1}{2} \Tr\{ (\Delta_k-\Delta_{k|k}) V_{k|k}^{-1} (\Delta_k-\Delta_{k|k})^\t R_k^{-1} \} \nonumber\\
&- \frac{\nu_{k|k}}{2} \log\det(R_k) - \frac{1}{2} \Tr\{ \Psi_{k|k} R_k^{-1} \}
\end{align}
where
\begin{align}
V_{k|k} =& \big( U_{k|k} + (u_{k|k}-\sqrt{\tfrac{2}{\pi}}\ones)(u_{k|k}-\sqrt{\tfrac{2}{\pi}}\ones)^\t + V_{k|k-1}^{-1} \big)^{-1} , \\
\Delta_{k|k} =& \big( (\ay_{k}-C_k x_{k|k})(u_{k|k}-\sqrt{\tfrac{2}{\pi}}\ones)^\t -C_k \Upsilon_{k|k} + \Delta_{k|k-1} V_{k|k-1}^{-1} \big) V_{k|k} , \\
\nu_{k|k} =& \nu_{k|k-1}+1, \\
\Psi_{k|k} =& \Delta_{k|k-1} V_{k|k-1}^{-1} \Delta_{k|k-1}^\t - \Delta_{k|k} V_{k|k}^{-1} \Delta_{k|k}^\t
+ (\ay_{k}-C_k x_{k|k})(\ay_{k}-C_k x_{k|k})^\t \nonumber\\
&+ C_k P_{k|k}C_k^\t + \Psi_{k|k-1} .
\end{align}
Therefore,
\begin{align}
q_{R,\Delta}(R_k,\Delta_k) &= \MN(\Delta_k; \Delta_{k|k}, R_k \kron V_{k|k}) \, \IW(R_k; \Psi_{k|k},\nu_{k|k}).
\end{align}
The following moments are required for the derivations of Section \ref{sec:Filterq_x}:
\begin{align}
&\E_{q_{R,\Delta}}[\Delta_k] = \Delta_{k|k},\\
&R_{k|k} \definedby \E_{q_{R,\Delta}}[R_k^{-1}]^{-1} = \frac{1}{\nu_{k|k}\!-\!n_z\!-\!1} \Psi_{k|k}. \label{eq:Rder}
\end{align}
Eq.\ \eqref{eq:Rder} follows from the fact that $R_k\!\sim\!\IW(\Psi_{k|k},\nu_{k|k})$ implies that $R_k^{-1}$ is Wishart-distributed with shape matrix $\Psi_{k|k}^{-1}$ and $\nu_{k|k}\!-\!n_z\!-\!1$ degrees of freedom \cite[Ch.\ 3.4]{GuptaN:2000}. Furthermore,
\begin{align}
&\E_{q_{R,\Delta}}[R_k^{-1} \Delta_k] \nonumber\\
&= \iint R_k^{-1} \Delta_k \, \MN(\Delta_k; \Delta_{k|k},R_k \kron V_{k|k}) \,  \IW(R_k; \Psi_{k|k},\nu_{k|k}) \d\Delta_k \d R_k \\
&= \int R_k^{-1} \Delta_{k|k} \, \IW(R_k; \Psi_{k|k},\nu_{k|k}) \d R_k \\
&= (\nu_{k|k}-n_z-1) \Psi_{k|k}^{-1} \Delta_{k|k} \\
&= R_{k|k}^{-1} \Delta_{k|k} 
\end{align}
and
\begin{align}
&\E_{q_{R,\Delta}}[\Delta_k^\t R_k^{-1} \Delta_k] \nonumber\\
&= \iint \Delta_k^\t R_k^{-1} \Delta_k \, \MN(\Delta_k; \Delta_{k|k},R_k \kron V_{k|k}) \,  \IW(R_k; \Psi_{k|k},\nu_{k|k}) \d\Delta_k \d R_k \\
&= \int ( \Tr\{R_k R_k^{-1}\} V_{k|k} + \Delta_{k|k}^\t R_k^{-1} \Delta_{k|k}) \, \IW(R_k; \Psi_{k|k},\nu_{k|k}) \d R_k \label{eq:EXTAX} \\
&= n_z V_{k|k} + (\nu_{k|k}-n_z-1) \Delta_{k|k}^\t \Psi_{k|k}^{-1} \Delta_{k|k} \\
&= n_z V_{k|k} + \Delta_{k|k}^\t R_{k|k}^{-1} \Delta_{k|k} ,
\end{align}
where \eqref{eq:EXTAX} follows from the matrix-variate normal identity $\E[X^\t A X] = \Tr\{UA^\t\}V+M^\t A M$ for $X\!\sim\!\N(M,U\kron V)$ \cite[Ch. 2.3]{GuptaN:2000}.

%% file: algo.tex

\section{Recursive Identification Algorithm for Linear Systems with Skewed Innovations}\label{table:filter}
\begin{algorithmic}[1]
\State \textbf{Inputs:} $x_{1|0}$,  $P_{1|0}$, $\Delta_{1|0}$, $V_{1|0}$, $\Psi_{1|0}$, $\nu_{1|0}$, $Q_{1:K}$, $C_{1:K}$, $\ay_{1:K}$, $\gamma$
\For{$k$ = 1 to $K$}
\Statex \hspace{3mm}\textit{Initialize}
\State $x_{k|k} \gets x_{k|k-1}$
\State $u_{k|k} \gets u_{k|k-1}$
\State $\Delta_{k|k} \gets \Delta_{k|k-1}$
\State $V_{k|k} \gets V_{k|k-1}$
\State $\Psi_{k|k} \gets \Psi_{k|k-1}$
\State $\nu_{k|k} \gets \nu_{k|k-1}+1$
\Repeat
	%
	\State $R_{k|k} \gets \tfrac{1}{\nu_{k|k}-n_z-1} \Psi_{k|k}$
\Statex \hspace{6mm}\textit{Update $q_{x,u}(x_{k},u_k) \approx \N(\left[ \begin{smallmatrix} x_k\\u_k \end{smallmatrix}\right]; \xi_{k|k},\Xi_{k|k})$}
	\State $\xi_{k|k-1} \gets \left[ \begin{smallmatrix} x_{k|k-1}\\n_z \sqrt{2/\pi}(\eye_{n_z}+n_z V_{k|k})^{-1} V_{k|k} \ones \end{smallmatrix} \right]$
	\State $\Xi_{k|k-1} \gets \mathrm{blockdiag}(P_{k|k-1}, (\eye+n_z V_{k|k})^{-1})$
	\State $\widetilde{C}_k \gets \begin{bmatrix} C_k&\Delta_{k|k} \end{bmatrix}$
	\State $K_k \gets \Xi_{k|k-1} \widetilde{C}_k^\t (\widetilde{C}_k \Xi_{k|k-1} \widetilde{C}_k^\t + R_{k|k})^{-1}$
	\State $\widehat\xi_{k|k}\gets \xi_{k|k-1}+K_k (\ay_{k}-\widetilde{C}_k \xi_{k|k-1}+\Delta_{k|k} \sqrt{\tfrac{2}{\pi}} \ones )$
	\State $\widehat\Xi_{k|k}\gets \Xi_{k|k-1} - K_k \widetilde{C}_k K_k^\t$
	\State $[\xi_{k|k}, \Xi_{k|k}] \gets \texttt{seq\_trunc}(\widehat\xi_{k|k}, \widehat\Xi_{k|k}, \{\nar\!+\!1,\ldots,\nar\!+\!n_z\})$ \hspace{-1.5mm} \Comment{See \cite[Table I]{nurminen2016b_arxiv}}
	\State $x_{k|k} \gets [\xi_{k|k}]_{1:\nar}$
	\State $P_{k|k} \gets [\Xi_{k|k}]_{1:\nar,1:\nar}$
	\State $\widetilde{u}_{k|k} \gets [\xi_{k|k}]_{\nar+(1:n_z)}\!-\!\sqrt{\tfrac{2}{\pi}}\ones$
	\State $U_{k|k} \gets [\Xi_{k|k}]_{\nar+(1:n_z),\nar+(1:n_z)}$
	\State $\Upsilon_{k|k} \gets [\Xi_{k|k}]_{1:\nar,\nar+(1:n_z)}$
	\Statex \hspace{6mm}\textit{Update $q_{R,\Delta}(R_k,\Delta_k)=\MN(\Delta_k; \Delta_{k|k}, R_k \kron V_{k|k})\,\IW(R_k; R_{k|k},\nu_{k|k})$ }
	\State $V_{k|k} \gets \big(U_{k|k}+\widetilde{u}_{k|k} \widetilde{u}_{k|k}^\t+V_{k|k-1}^{-1} \big)^{-1}$
	\State $\Delta_{k|k} \gets \big((z_{k}-C_k x_{k|k}) \widetilde{u}_{k|k}^\t - C_k \Upsilon_{k|k} + \Delta_{k|k-1}V_{k|k-1}^{-1} \big) V_{k|k}$
	\State $\Psi_{k|k} \gets \Delta_{k|k-1} V_{k|k-1}^{-1} \Delta_{k|k-1}^\t - \Delta_{k|k} V_{k|k}^{-1} \Delta_{k|k}^\t$
	\State \hspace{3mm} $+ (z_{k}-C_k x_{k|k})(z_{k}-C_k x_{k|k})^\t + C_k P_{k|k}C_k^\t + \Psi_{k|k-1}$
\Until{\textbf{converged}}
\Statex \hspace{3mm}\textit{Predict}
\State $x_{k+1|k} \gets x_{k|k}$
\State $P_{k+1|k} \gets P_{k|k}+Q_k$
\State $\Delta_{k+1|k} \gets \Delta_{k|k}$
\State $V_{k+1|k} \gets \frac{1}{\gamma} V_{k|k}$
\State $\Psi_{k+1|k} \gets \gamma \Psi_{k|k}$
\State $\nu_{k+1|k} \gets \gamma\, \nu_{k|k} + (1\!-\!\gamma)\! \cdot\! 2n_z$
\EndFor
\State \textbf{Outputs: $x_{k|k}$ and  $P_{k|k}$ for $k=1,\ldots,K$ } 
\end{algorithmic}

%% file: nips_2016.bbl
\begin{thebibliography}{10}

\bibitem{harvey1999}
C.~R. Harvey and A.~Siddique, ``Autoregressive conditional skewness,'' {\em The
  Journal of Financial and Quantitative Analysis}, vol.~34, pp.~465--487,
  December 1999.

\bibitem{jondeau2003}
E.~Jondeau and M.~Rockinger, ``Conditional volatility, skewness, and kurtosis:
  existence, persistence, and comovements,'' {\em Journal of Economic Dynamics
  and Control}, vol.~27, pp.~1699--1737, 2003.

\bibitem{christofferssen2006}
P.~Christofferssen, S.~Heston, and K.~Jacobs, ``Option valuation with
  conditional skewness,'' {\em Journal of Econometrics}, vol.~131,
  pp.~253--284, 2006.

\bibitem{tsiotas2012}
G.~Tsiotas, ``On generalised asymmetric stochastic volatility models,'' {\em
  Computational Statistics and Data Analysis}, vol.~56, pp.~151--172, 2012.

\bibitem{kaemarungsi2012}
K.~Kaemarungsi and P.~Krishnamurthy, ``Analysis of {WLAN}'s received signal
  strength indication for indoor location fingerprinting,'' {\em Pervasive and
  Mobile Computing}, vol.~8, no.~2, pp.~292--316, 2012.
\newblock Special Issue: Wide-Scale Vehicular Sensor Networks and Mobile
  Sensing.

\bibitem{nurminen2015b}
H.~Nurminen, T.~Ardeshiri, R.~Pich\'{e}, and F.~Gustafsson, ``A {NLOS}-robust
  {TOA} positioning filter based on a skew-$t$ measurement noise model,'' in
  {\em International Conference on Indoor Positioning and Indoor Navigation
  (IPIN)}, pp.~1--7, October 2015.

\bibitem{azzalini1985}
A.~Azzalini, ``A class of distributions which includes the normal ones,'' {\em
  Scandinavian Journal of Statistics}, vol.~12, no.~2, pp.~171--178, 1985.

\bibitem{azzalini1996}
A.~Azzalini and A.~Dalla~Valle, ``The multivariate skew-normal distribution,''
  {\em Biometrika}, vol.~83, no.~4, pp.~715--726, 1996.

\bibitem{lin2009}
T.~I. Lin, ``Maximum likelihood estimation for multivariate skew normal mixture
  models,'' {\em Journal of Multivariate Analysis}, vol.~100, pp.~257--265,
  2009.

\bibitem{lee2014}
S.~Lee and G.~J. McLachlan, ``Finite mixtures of multivariate skew
  t-distributions: some recent and new results,'' {\em Statistics and
  Computing}, vol.~24, no.~2, pp.~181--202, 2014.

\bibitem{ho2012}
H.~J. Ho, S.~Pyne, and T.~I. Lin, ``Maximum likelihood inference for mixtures
  of skew student-$t$-normal distributions through practical {EM}-type
  algorithms,'' {\em Statistics and Computing}, vol.~22, pp.~287--299, 2012.

\bibitem{nurminen2015a}
H.~Nurminen, T.~Ardeshiri, R.~Pich\'e, and F.~Gustafsson, ``Robust inference
  for state-space models with skewed measurement noise,'' {\em IEEE Signal
  Processing Letters}, vol.~22, pp.~1898--1902, November 2015.

\bibitem{nurminen2016b_arxiv}
H.~Nurminen, T.~Ardeshiri, R.~Pich\'{e}, and F.~Gustafsson, ``Skew-$t$ filter
  and smoother with improved covariance matrix approximation.'' Available
  online at \url{http://arxiv.org/abs/1608.07435}, August 2016.

\bibitem{ljung2002}
L.~Ljung, ``Recursive identification algorithms,'' {\em Circuits, Systems, and
  Signal Processing}, vol.~21, no.~1, pp.~57--68, 2002.

\bibitem{arellanovalle2005}
R.~B. Arellano-Valle and M.~G. Genton, ``On fundamental skew distributions,''
  {\em Journal of Multivariate Analysis}, no.~96, pp.~93--116, 2005.

\bibitem{lee2016}
S.~X. Lee and G.~J. McLachlan, ``Finite mixtures of canonical fundamental skew
  $t$-distributions -- the unification of the restricted and unrestricted skew
  $t$-mixture models,'' {\em Statistics and {C}omputing}, no.~26, pp.~573--589,
  2016.

\bibitem{GuptaN:2000}
A.~K. Gupta and D.~K. Nagar, {\em Matrix variate distributions}.
\newblock Boca Raton, FL: Chapman \& Hall/CRC, 2000.

\bibitem{Bishop2007}
C.~M. Bishop, {\em Pattern Recognition and Machine Learning}.
\newblock Springer, 2007.

\bibitem{TzikasLG2008}
D.~G. Tzikas, A.~C. Likas, and N.~P. Galatsanos, ``The variational
  approximation for {B}ayesian inference,'' {\em IEEE Signal Processing
  Magazine}, vol.~25, pp.~131--146, Nov. 2008.

\bibitem{perala2008}
T.~Per\"{a}l\"{a} and S.~Ali-L\"{o}ytty, ``{Kalman-type positioning filters
  with floor plan information},'' in {\em 6th International Conference on
  Advances in Mobile Computing and Multimedia (MoMM)}, pp.~350--355, 2008.

\bibitem{simon2010}
D.~J. Simon and D.~L. Simon, ``Constrained {K}alman filtering via density
  function truncation for turbofan engine health estimation,'' {\em
  International Journal of Systems Science}, vol.~41, no.~2, pp.~159--171,
  2010.

\bibitem{karny2012}
M.~K\'arn\'y and K.~Dedecius, ``Approximate {B}ayesian recursive estimation: On
  approximation errors,'' tech. rep., \'UTIA AV \v{C}R, January 2012.

\bibitem{ozkan2013}
E.~\"Ozkan, V.~\v{S}m\'idl, S.~Saha, C.~Lundquist, and F.~Gustafsson,
  ``Marginalized adaptive particle filtering for nonlinear models with unknown
  time-varying noise parameters,'' {\em Automatica}, vol.~49, 2013.

\bibitem{pillonetto2010}
G.~Pillonetto and G.~De~Nicolao, ``A new kernel-based approach for linear
  system identification,'' {\em Automatica}, vol.~46, pp.~81--93, 2010.

\bibitem{chen2012}
T.~Chen, H.~Ohlsson, and L.~Ljung, ``On the estimation of transfer functions,
  regularizations and {G}aussian processes--revisited,'' {\em Automatica},
  vol.~48, pp.~1525--1535, 2012.

\bibitem{Agamennoni12}
G.~Agamennoni, J.~Nieto, and E.~Nebot, ``Approximate inference in state-space
  models with heavy-tailed noise,'' {\em IEEE Transactions on Signal
  Processing}, vol.~60, pp.~5024--5037, October 2012.

\end{thebibliography}
